\documentclass[twoside]{article}
\usepackage{fleqn,espcrc2}

\usepackage{epsfig}

\newcommand{\beeq}{\begin{equation}}
\newcommand{\beeqar}{\begin{eqnarray}}
\newcommand{\eneq}{\end{equation}}
\newcommand{\eneqar}{\end{eqnarray}}

\newcommand{\AmS}{{\protect\the\textfont2
  A\kern-.1667em\lower.5ex\hbox{M}\kern-.125emS}}

\title{ Quasi-local Update Algorithms for Numerical Simulations of d=3
SU(2) LGT in the Dual Formulation }

\author{N.D. Hari Dass\address[IMSC]{Institute of Mathematical Sciences, 
        C.I.T Campus, Chennai 600 113, INDIA\\ }%
        \thanks{{\bf email: dass@imsc.ernet.in}}}
       
\begin{document}

\begin{abstract}
In the dual formulation of d=3 SU(2) LGT, the link variables
are group representations and valid configurations are those
satisfying a number of triangle inequalities. In \cite{lat99}
algorithms for local updates that automatically respect these
constraints were described. It was also pointed out there that these
local updates were not ergodic. In this presentation, we describe
two different quasi-local updating algorithms which, in conjunction
with the local updates, appear to be ergodic.
\vspace{1pc}
\end{abstract}

\maketitle

\section{THE DUAL FORMULATION}
A brief introduction to the dual formalism
for $d=3$ $SU(2)$ lattice gauge theory 
as well as to the techniques that have been developed for its numerical
simulations was presented at LATTICE'99 \cite {lat99}. As was stated there, 
the partition function of the conventional LGT can be converted,
upon using the techniques of character expansion and group integrations,
into the dual form \cite{anishetty} 
\beeq
Z_{d} =\sum_{\{j\}}\prod (2j+1)C_{j_a}(\beta)
\prod_{i=1}^{5}\left\{\matrix{a_i&b_i&c_i\cr d_i&e_i&f_i\cr }\right\}
\eneq
This partition sum is defined over the dual lattice where $\{j_a\}$ live over
the links and $\{j_b\}$ over the diagonals to the plaquettes. The convention for
the diagonals is that they connect the vertices of the odd sublattice. We have
collectively designated $\{j_a\},\{j_b\}$ by $\{j\}$. Each cube of the dual lattice
is spanned by 5 tetrahedra of which one is spanned entirely by $\{j_b\}$
while four are spanned by three $\{j_a\}$ and three $\{j_b\}$. 
In what follows, the b-links will be shown by dashed lines while the a-links
will be shown by solid lines.
Each tetrahedron
carries a weight \mbox{factor which is the $SU(2)$ 6-j symbol $\left\{\matrix{a&b&c\cr
d&e&f\cr}\right\}$.} \mbox{Periodic b.c for the original lattice is} 
crucial for this construction.
\section{QUASI-LOCAL UPDATES}
An important ingredient in the techniques for numerical
simulation of such classes of models is an updating
technique that respects all the triangle inequalities. In
\cite{lat99} we introduced two distinct classes of
updating algorithms: {\bf local} and {\bf quasi-local}.
The local algorithms work very much like the usual
updating techniques for Monte Carlo simulations: all links
but one are held fixed and the free link is updated by
either Metropolis or heat bath methods. In the present
context, one can see immediately that such local
updates are not ergodic. To see this, let us recall that
the representations of $SU(2)$ can be grouped into two
classes i.e {\it half-integral} which can be called {\it
fermionic} and {integral} which can be called {\it
bosonic}. Let us call a move that takes fermionic links
to bosonic, and vice versa, a $Z_2$-flip. It is obvious
that in every triangle exactly two links must be
$Z_2$-flipped simulateneously if we are to maintain the
triangle inequalities. As the local moves change only one
link at a time, they can not accommadate $Z_2$-flips
and the moves therefore never take configurations out of
their $Z_2$-classes.

To circumvent this the {\bf quasi-local} moves were
introduced. Since in each triangle exactly two
$Z_2$-flips have to be carried out, these flips can
proliferate throughout the lattice unless a way out is 
found. Therefore one can look for those clusters of
$Z_2$-flips that have the smallest volume possible.
In this way we identified the {\bf globeven} moves at
even-sublattice points which flip the six interior
a-links as shown in fig.1. All the "exterior" b-links
are held fixed during a globeven move and this contains
the $Z_2$-flips to the interior. It should be noted
that eight tetrahedra get updated during each globeven
move.

\begin{figure}[htb]
\begin{center}
\mbox{\epsfig{file=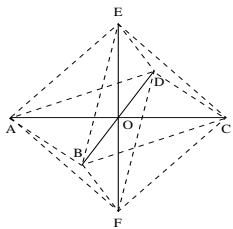,width=3truecm,height=3truecm,angle=-90}}
\caption{Quasi-local update at even sites. }
\label{Fig 1.}
\end{center}
\end{figure}

While the {\bf globeven} moves are capable of
$Z_2$-flipping
the a-links, they leave the b-links unchanged. For this
reason, these moves are also not ergodic. Since b-links
connect only odd sites, any quasi-local move affecting
them must take place at the odd sites. It then follows
that the smallest volume of $Z_2$-flippings involves
all the 6 a-links and 12 b-links at the odd site while
keeping all the other a-links and b-links fixed with 32
tetrahedra getting updated. This is
illustrated in fig.2.

\begin{figure}[htb]
\begin{center}
\mbox{\epsfig{file=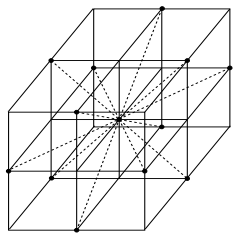,width=3truecm,height=3truecm,angle=-90}}
\caption{Quasi-local updates at odd sites.}
\label{Fig 2.}
\end{center}
\end{figure}

\section{KAGOME VARIABLES}
Though we have identified the geometry of the cluster of links
to be updated during quasi-local moves, we have to address the issue of
how the new spin values are to be chosen such that all triangle
inequalities are respected. Let us first recall the form of triangle
inequalities. Considering the triangle $(j_1,j_2,j_{12})$ of fig.3,
these inequalities take the form
\beeqar
|j_1-j_2|~~~\le&~j_{12}~~\le&~j_1+j_2\nonumber\\
|j_1-j_{12}|~~~\le&~j_{2}~~\le&~j_1+j_{12}\nonumber\\
|j_2-j_{12}|~~~\le&~j_{1}~~\le&~j_2+j_{12}
\eneqar
\begin{figure}[htb]
\begin{center}
\mbox{\epsfig{file=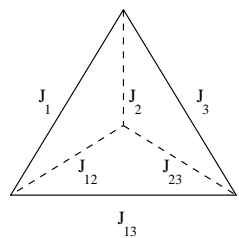,width=3truecm,height=3truecm,angle=-90}}
\caption{A Single Tetrahedron.}
\label{Fig 3.}
\end{center}
\end{figure}

We illustrate the K-variable strategy with the example of a single
tetrahedra. The idea behind these variables is to replace inequalities
by equalities to some extent. They were first introduced by Bargman \cite{barg}
and were later developed by Anishetty et al \cite{gadiyar}. Again, let us first
introduce them for the triangle considered above. Their construction
involves a pair $(n_1,n_2)$ for each link or $j$-value such that
\beeqar
n_1+n_2~&= j_1\nonumber\\
n_1+n_3~&= j_2\nonumber\\
n_3+n_2~&= j_{12}
\eneqar
These relations are invertible:
\beeqar
2n_1~&={j_1+j_2-j_{12}}\nonumber\\
2n_2~&={j_{12}+j_1-j_2}\nonumber\\
2n_3~&={j_2+j_{12}-j_1}
\eneqar
From eqn (4) it is clear that $n_i$ are either integers or half-integers
and also positive semi-definite. The moral of the story is that we can
choose any triplet $(n_1,n_2,n_3)$ and the $j_i$ constructed from them
according to eqn (3) will automatically satisfy the triangle
inequalities of eqn (2).

Now let us consider the tetrahedron of fig.3 and ask how we can choose
$(j_1,j_2,j_3,j_{12},j_{23},j_{13})$ such that the four sets of triangle
inequalities are all satisfied. At the level of $j_i$ this can be done
as follows: freely choose 3 $j$'s originating at a vertex e.g
$(j_1,j_2,j_3)$. Next choose $j_{12}$ in the range given in eqn (2)
and choose $j_{23}$ likewise. Finally choose $j_{13}$ such that it
simultaneously satisfies the triangle inequalities relevant for the
two triangles $(j_1,j_3,j_{13})$ and $(j_{12},j_{23},j_{13})$. This
procedure gets increasingly tedious.
Not only is this procedure tedious, it does not even guarantee a
solution always even though in the case of the single tetrahedron a
solution is always possible. 

That a solution using this strategy may not always be possible can be
seen by applying it to the gobeven geometry of fig.1. Here one could
start by first selecting a $Z_2$-flipped value for, say, OE. The link
OA could then be chosen to satisfy the inequalities for OAE, the link
OB chosen to satisfy OBE and OAB. So far the situation is like that for
a single tetrahedron and all would be fine. Next, OC will have to be
chosen to satisfy OCE and OBC; already at this stage there is no
guarantee that these two sets can be simultaneously realised. OD will
have to be chosen to satisfy ODE, AOD and COD; OF will have to satisfy
FOA, FOB, FOC and FOD! The situation only gets worse with these additional
requirements.

Now we show how the same problem can be efficiently addressed by using
the K-variables.Again we illustrate the method by first applying it to
the simple case of a single tetrahedron. For this purpose we introduce 
the variables
$(n_1,n_2,n_3)$ for the set $(j_1,j_2,j_{12})$, $(n_4,n_5,n_6)$ for
$(j_2,j_3,j_{23})$, $(n_7,n_8,n_9)$ for $(j_1,j_3,j_{13})$, and 
$(l_1,l_2,l_3)$ for $(j_{12},j_{23},j_{13})$. 
Writing down the various equalities we get
\beeqar
j_1~&=~~n_2+n_3~&=~~n_8+n_9\nonumber\\
j_2~&=~~n_1+n_3~&=~~n_5+n_6\nonumber\\
j_3~&=~~n_4+n_6~&=~~n_7+n_9
\eneqar
and
\beeqar
j_{12}~&=~~n_1+n_2~&=l_2+l_3\nonumber\\
j_{23}~&=~~n_4+n_5~&=l_1+l_3\nonumber\\
j_{13}~&=~~n_7+n_8~&=l_2+l_1
\eneqar
From eqn(5) it follows that $(n_1,n_2,n_3,n_4)$ can be chosen to be
unrestricted, one of $(n_5,n_6)$ can be chosen to be independent but
with restricted range and one of $(n_7,n_8,n_9)$ to be also independent
but with restricted range. Choosing $(n_6,n_9)$ to be the independent
ones, the relevant restricted ranges are
\beeqar
0~~\le&~~n_6~~\le&~~n_1+n_3\nonumber\\
0~~\le&~~n_9~~\le&~~min(n_2+n_3,n_4+n_6)
\eneqar
Though $(l_1,l_2,l_3)$ are determined by the $n_i$, the fact they all
have to be positive semidefinite yields additional restrictions which are
best seen by solving for $l_i$ using eqn (6):
\beeqar
l_1~~&=~~n_3+n_4-n_9\nonumber\\
l_2~~&=~~n_2+n_6-n_9\nonumber\\
l_3~~&=~~n_1+n_9-n_6
\eneqar
Summarising, the solution to the single tetrahedron problem in terms of
the K-variables is: $(n_1,n_2,n_3,n_4)$ are unrestricted except for
positive semi-definiteness, $(n_6,n_9)$ are independent but restricted
in range and $(n_5,n_7,n_8)$ are dependent:
\beeqar
0\le&n_6\le&n_1+n_3\nonumber\\
n_6-n_1\le&n_9\le&min(n_2,n_4)+min(n_3,n_6)
\eneqar
and
\beeqar
n_5& = n_1+n_3-n_6&\nonumber\\
n_7& = n_4+n_6-n_9&\nonumber\\
n_8& = n_2+n_3-n_9&
\eneqar
\subsection{TWO TETRAHEDRA CELL}
Our solution to the single tetrahedron given by eqn(9) will form the
basis for the eventual solution of the globeven and globodd problems.
The final solution will be arrived at in three stages: i) the solution
of the two-tetrahedra problem, ii) the solution of the four-tetrahedra
problem and iii) globeven and globodd using the results of ii). We now
show how the two-tetrahedra problem is solved. For this purpose, consider
two tetrahedra glued together as shown in fig.4. We take the common
triangle ABC with $AB = j_1,BC = j_2, CA = j_{12}$ to be associated with the
independent K-variables $(n_1,n_2,n_3)$. For the second tetrahedron ABCD
we have to introduce 6 more variables but there will be 3 more relations
associated with the links AB,BC,CA thus reducing the additional
variables to 3. 
\begin{figure}[htb]
\begin{center}
\mbox{\epsfig{file=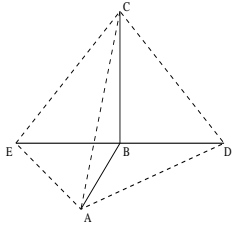,width=3truecm,height=3truecm,angle=-90}}
\caption{A two-tetrahedra cell.}
\label{Fig 4.}
\end{center}
\end{figure}

The spin assignments are taken to be : $ BE = j_3, AE =
j_{13}, CE = j_{23}$ and $ BD = k_3, AD = k_{13}, CD = k_{23}$ and the
additional K-variable assignments to be : $(n_4,n_5,n_6)$ with 
$(j_3,j_2,j_{23})$, $(n_7,n_8,n_9)$ with $(j_1,j_3,j_{13})$,
$(m_4,m_5,m_6)$ with $(k_3,j_2,k_{23})$, and $(m_7,m_8,m_9)$ with
$(j_1,k_3,k_{13})$. We give only the final results: the unrestricted
variables are $(n_1,n_2,n_3,n_5,m_5)$ and the restricted but independent
variables are $(n_6,n_9,m_6,m_9)$ and the remaining dependent variables
are $(n_4,n_7,n_8,m_4,m_7,m_8)$. The equation corresponding to eqn (9)
are:
\beeqar
0\le&n_6\le&n_1+n_3\nonumber\\
n_6-n_1\le&n_9\le&min(n_2,n_5)+min(n_3,n_6)\nonumber\\
0\le&m_6\le&n_1+n_3\nonumber\\
m_6-n_1\le&m_9\le&min(n_2,m_5)+min(n_3,m_6)\nonumber\\
\eneqar
and
\beeqar
n_4& = n_1+n_3-n_6\nonumber\\
n_7& = n_5+n_6-n_9\nonumber\\
n_8& = n_2+n_3-n_9\nonumber\\
m_4& = n_1+n_3-m_6\nonumber\\
m_7& = m_5+m_6-m_9\nonumber\\
m_8& = n_2+n_3-m_9
\eneqar
\subsection{FOUR TETRAHEDRA CELL}
As the final ingredient in the K-variable solution to the quasi-local
updates, we now consider four tetrahedra glued together as shown in
fig.5. In the two tetrahedra case we finally had 15 K-variables of which
9 were independent and 6 dependent. The K-variable assignments 
are summarised below:
\beeqar
(OA,AE,OE)~~~&\rightarrow~~~(n1,n2,n3)\nonumber\\
(DA,AE,DE)~~~&\rightarrow~~~(n4,n5,n6)\nonumber\\
(OA,AD,OD)~~~&\rightarrow~~~(n7,n8,n9)\nonumber\\
(OA,AC,OC)~~~&\rightarrow~~~(p1,p2,p3)\nonumber\\
(DA,AC,DC)~~~&\rightarrow~~~(p4,p5,p6)\nonumber\\
(BA,AC,BC)~~~&\rightarrow~~~(q4,q5,q6)\nonumber\\
(BA,AE,BE)~~~&\rightarrow~~~(m4,m5,m6)\nonumber\\
(OA,AB,OB)~~~&\rightarrow~~~(m7,m8,m9)
\eneqar
\begin{figure}[htb]
\begin{center}
\mbox{\epsfig{file=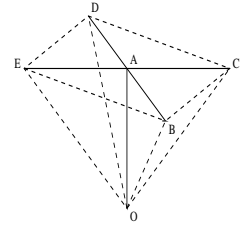,width=3truecm,height=3truecm,angle=-90}}
\caption{A four-tetrahedra cell. }
\label{Fig 5.}
\end{center}
\end{figure}
Of the 24 K-variables 6 are unrestricted, 7 are restricted but
independent and 11 are dependent. These are summarised in the
following:
\beeqar
n_1+n_3 \ge&n_6\ge& 0 \nonumber\\
n_1+n_3 \ge&m_6\ge& 0 \nonumber\\
n_2+n_3 \ge&p_2\ge& 0 
\eneqar
\beeqar
min(n_2,n_5)+min(n_3,n_6) \ge&n_9\ge& n_6-n_1 \nonumber\\
min(n_2,m_5)+min(n_3,m_6) \ge&m_9\ge& m_6-n_1 \nonumber
\eneqar
\beeqar
min(n_5+n_6,n_2+n_3+p_1-p_2) \ge&p_6\ge& 0 \nonumber\\
min(m_5+m_6,n_2+n_3+p_1-p_2) \ge&q_6\ge& 0\nonumber\\
\eneqar
The new feature at this level is that range restrictions are now of two
types: those that depend only on the unrestricted set as in the case of
$(n_6,m_6,p_2)$ and those that depend on the already restricted set as
in the case of $(n_9,m_9,p_6,q_6)$. The dependent variables are given
by:
\beeqar
n_4& = n_1+n_3-n_6\nonumber\\
n_7& = n_5+n_6-n_9\nonumber\\
n_8& = n_2+n_3-n_9\nonumber\\
m_4& = n_1+n_3-m_6\nonumber\\
m_7& = m_5+m_6-m_9 \nonumber\\
m_8& = n_2+n_3-m_9 \nonumber\\
p_3& = n_2+n_3-p_2 \nonumber\\
p_4& = p_1+p_3-p_6 \nonumber\\
p_5& = n_5+n_6-p_6 \nonumber\\
q_4& = p_1+p_3-q_6 \nonumber\\
q_5& = m_5+m_6-q_6 
\eneqar
\section{GLOBEVEN AND GLOBODD}
Finally we are in a position to solve the globeven and globodd problems.
For globeven, one takes two copies of the 4-tetrahedra cell and glues
them such that the rectangular bases of the two cells coincide. As before
the number of K-variables required will be twice the number for each
4-tetrahedra cell minus the number associated with the rectangular base.
But because of the 8 links that are shared between the two cells, the
independent variables will be reduced by 8. The results are
similar to eqns(15,16) but more lengthy and will not be displayed here.

For the globodd case we have to take six copies of the 4-tetrahedra
cells and glue them together such that the six rectangular bases now
form the skin of the 8-cube cluster shown in fig.2. Now the total number
of independent K-variables will be six times those of the 4-tetrahedra
cell reduced by the 12 new relations that arise due to the sharing of
the 12 b-links.

This completes the solution to the problem of finding
spin-configurations for the globoeven and globodd updates
that will respect all the triangle inequalities. Yet, the
resulting equations are quite involved and may not result in
efficient algorithms for numerical simulations. A practical
approach that can be followed to accomplish a $Z_2$-flip is 
to simply increase all the interior spins by ${1\over 2}$
except those that are already at the maximum value permitted by the
code; for these, the spin value can be reduced by ${1\over 2}$. It is
easily seen that this "brute force" $Z_2$-flip will also maintain the
triangle inequalities. But such moves do not necessarily respect
the balance conditions required of Monte Carlo simulations. As a
way out of this, we swept through the relevant clusters many times
randomly setting the links to new values allowed by triangle inequalities.
This was achieved through the local update methods mentioned
before. New values were accepted according to a metropolis
algorithm or a heat bath method carried over a small chosen subset. 
The random numbers were generated uniformly in the range
$(0,1)$..Whether this method is picking new link samples in an unbiased
way is yet to be understood satisfactorily \cite{lat00}.

\end{document}